\documentclass[twocolumn,showpacs,preprintnumbers,prl]{revtex4}
           \usepackage{epsfig}
           \usepackage{graphics}
           \usepackage{amsmath}
           \usepackage{color}
             \definecolor{gray}{rgb}{0.8,0.8,0.8}
             \definecolor{darkgreen}{rgb}{0.0,0.6,0.0}
           \usepackage{bm}
           \usepackage{dcolumn}
\begin{document}

\title{%
  Magnetic proximity effect in Perovskite Superconductor/Ferromagnet Multilayers
}

\author{%
  J.\,Stahn$^1$, J.\,Chakhalian$^2$, Ch.\,Niedermayer$^1$, J.\,Hoppler$^1$,
  T.\,Gutberlet$^1$, J.\,Voigt$^1$, F.\,Treubel$^3$, H-U.\,Habermeier$^2$,
  G.\,Cristiani$^2$, B.\,Keimer$^2$ and
  C.\,Bernhard$^2$
}

\affiliation{%
  $^1$Laboratorium f\"ur Neutronenstreuung, ETH Z\"urich \& PSI, Villigen,
  Switzerland
}
\affiliation{%
  $^2$Max Planck Institut f\"ur Festk\"orperforschung, Stuttgart, Germany
}
\affiliation{%
  $^3$Fakult\"at f\"ur Physik, Universit\"at Konstanz, Germany
}

\date{%
  \today
}

\begin{abstract}%
$\mathrm{YBa_2Cu_3O_7/La_{2/3}Ca_{1/3}MnO_3}$ 
 superconducting/ferromagnetic (SC/FM)
multilayers have been studied by neutron reflectometry. Evidence for a
characteristic difference between the structural and magnetic
depth profiles is obtained from the occurrence of a structurally
forbidden Bragg peak in the FM state.
The comparison with simulated reflectivity curves allows us to identify two
possible magnetization profiles: a sizable magnetic moment within
the SC layer antiparallel to the one in the FM layer (inverse
proximity effect), or a ``dead'' region in the FM layer with zero
net magnetic moment. The former scenario is supported by an
anomalous SC-induced enhancement of the off-specular reflection,
which testifies to a strong mutual interaction of SC and FM order
parameters.
\end{abstract}


\maketitle
%%%%%%%%%%%%%%%%%%%%%%%%%%%%%%%%%%%%%%%%%%%%%%%%%%%%%%%%%%%%%%%%%%%%%%%%%%%%%%

Recent advances in fabrication and characterization of multilayers
with nanoscale periodicity based on perovskite oxides have opened
a new avenue in the investigation of materials with strong
electron correlations \cite{chen}. Superlattices composed of
ferromagnets (FM) and superconductors (SC) are of particular
interest because their mutually exclusive ground state properties
can give rise to novel quantum phenomena \cite{izyu}. Prominent
examples are the so-called $\pi$-junction effect \cite{rosa},
where the phase of the SC order parameter is modulated across the
layers, or states with a spatial modulation of the amplitudes of
the FM and SC order parameters such as spontaneous vortex phases
or the Larkin-Ovchinikov-Fulde-Ferrel (LOFF) state. Experimental
signatures include a non-trivial dependence of $T_\text{c}$ on the
FM layer thickness \cite{sefr,habe} and a complex magnetic phase
diagram with reentrant SC states.

Extensive earlier work on classical metallic FM/SC multilayers has
verified the $\pi -$junction effect, whereas a LOFF pairing state
has not yet been firmly established. The work on the perovskite
oxide FM/SC superlattices is motivated by the appealing properties
of the cuprate high $T_\text{c}$ superconductors (HTSC) whose high
SC critical temperatures make them potentially useful for
technological applications. Further, since HTSC are believed to be
susceptible to a variety of competing instabilities, there is a
high potential for novel SC/FM quantum states in multilayer
structures. This research is in its early stage, and relatively
little is known about the nature of magnetism at the interface,
the spatial distribution of the magnetization throughout the
layers, and the interplay of FM and SC order parameters in
general. Neutron reflectivity has been a tool of choice in
investigating interfaces in thin films and multilayers
\cite{felc,edxr}. In general, it allows one to probe a potential
normal to the surface which consists of the contributions from the
atomic nuclei $V_\text{nuc}(z)$ and the magnetic potential
$V_{\text{mag}}(z)$. A periodic multilayer can be regarded as a
one-dimensional crystal which gives rise to Bragg peaks that
provide information on the number of layers from the peak width,
the period length from the distance between adjacent peaks, and
the ratio of the individual layer thicknesses from the relative
peak intensities. In addition, the in-plane (``off-specular")
width of the peaks yields information about the in-plane
magnetization profile. This technique has provided valuable
information on the microscopic magnetic properties of classical
FM/SC multilayers, but has so far not been successfully applied to
perovskite oxide FM/SC multilayers.

In this letter we report the first results of polarized and
unpolarized neutron reflectivity measurements on symmetric
superlattices (with identical LCMO and YBCO layer thicknesses) 
that consist of alternating layers of the FM
colossal magneto-resistance material
$\mathrm{La_{2/3}Ca_{1/3}MnO_3}$ (LCMO) and the HTSC compound
$\mathrm{YBa_2Cu_3O_7}$ (YBCO). Symmetric superlattices  are well suited to
explore a possible interference between SC and FM order
parameters, because an extinction rule disallows all even-order
Bragg reflections if $V_{\text{mag}}(z)$ is spatially uniform and
confined to the LCMO layer. The reflectivity curves above the FM
and SC transitions ($T > T_\text{mag}$ and $T_\text{sc}$) indeed
exhibit only odd-numbered Bragg peaks and testify to the
high structural quality of our superlattices (with an rms
interface roughness \cite{edxr} of $\sigma \approx
5\,\text{\AA}$). The reflectivity curves exhibit marked changes
in the FM state as well as in the SC state. In particular, the
appearance of a second-order magnetic Bragg peak below
$T_\text{mag}$ indicates that  $V_{\text{mag}}(z)$ either reaches
into the YBCO layer with antiferromagnetic coupling across the interface 
or is confined to a spatial range
significantly less than the thickness of the LCMO layer. Both
scenarios are incompatible with a conventional magnetic proximity
effect as proposed in Refs. \cite{melo,rado}.
An anomalous enhancement of the off-specular reflection in the SC
state indicates a strong mutual interaction of SC and FM order
parameters. This lends support to the former model, where the SC
and FM order parameters are in intimate contact, and disfavors the
latter one, where they are separated by a magnetically ``dead''
region.

Superlattices of
$\mathrm{[LCMO(98\text{\,\AA})/YBCO(98\text{\,\AA})]_{7}}$ (sample
1) and
$\mathrm{[LCMO(160\text{\,\AA})/YBCO(160\text{\,\AA})]_{6}}$
(sample 2) were grown by pulsed laser deposition (PLD) on
$10\times 10\times 0.5\,\text{mm}^{3}$ $\mathrm{SrTiO_3}$ (001)
substrates \cite{habe}. Their high quality was confirmed by x-ray
diffraction, which showed epitaxial growth of the films with the
$c$-axis along (110). Resistivity and SQUID magnetization
measurements revealed a FM transition at $T_\text{mag} \approx
165\,\text{K}$ and the onset of SC at $T_\text{sc}\approx
75\,\text{K}$. These values are substantially reduced from the
typical bulk values of $T_\text{mag}^\text{LCMO} = 270\,\text{K}$
and $T_\text{sc}^\text{YBCO} = 93\,\text{K}$, likely due to the
proximity effect. \cite{sefr,habe}

Unpolarized and polarized angle-dispersive neutron reflectivity
measurements have been performed on the 2-axes diffractometers
Morpheus at SINQ and ADAM at ILL. The polarized neutrons were
reflected from the superlattice into a $^3$He detector. The
samples were mounted in a closed-cycle refrigerator with a
temperature range from 12 to 300\,K. The external magnetic field
$H_\text{ext}$ produced by Helmholtz coils was oriented
perpendicular to the scattering plane and parallel to the film
surface.

           %================================================
           \begin{figure}[t]
            \unitlength1mm
            %%%%%%%%%%%%%%%%%%%%%%%%%%%%%%%%%%%%%%%%%%%%%%%%
            \begin{picture}(80,55)
   \put(000,051){(a)}
   \put(041,008){{\color{gray}\rule{9mm}{29mm}}}
   \put(002,003){\includegraphics[width=77mm]{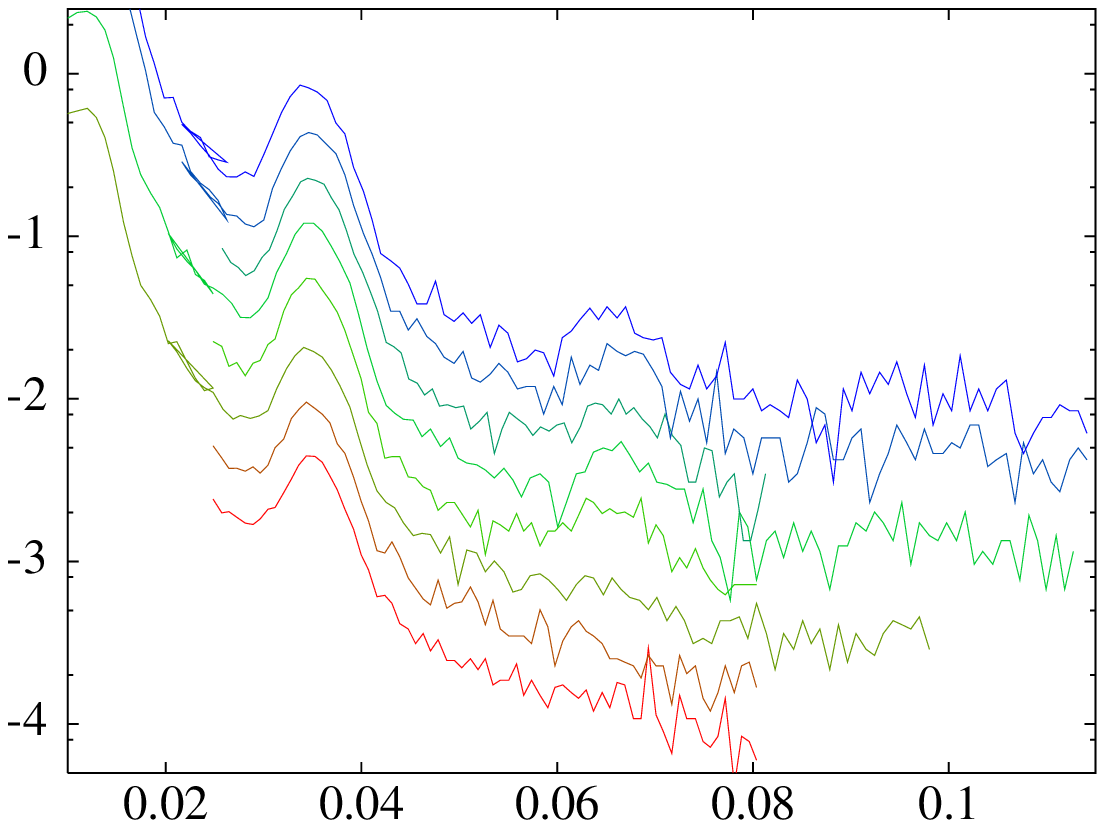}}
   \put(002,022){\small\rotatebox{90}{$\log_{10}[R(q_z)]$}}
   \put(040,000){\small $q_z$ (\AA$^{-1}$)}
   \put(027,051){$\downarrow  1^\text{st}$}
   \put(045,038){$\downarrow  2^\text{nd}$}
   \put(064,039){$\downarrow  3^\text{rd}$}
   \put(069,031){\color{blue}$15\,\text{K}$}
   \put(056,017){\color{green}$150\,\text{K}$}
   \put(056,008){\color{red}$200\,\text{K}$}
            \end{picture}
            %%%%%%%%%%%%%%%%%%%%%%%%%%%%%%%%%%%%%%%%%%%%%%%%
            \begin{picture}(80,32)
   \put(000,029){(b)}
   \put(002,000){%
      \includegraphics[bb= 90 80 327 245,clip,width=40mm]{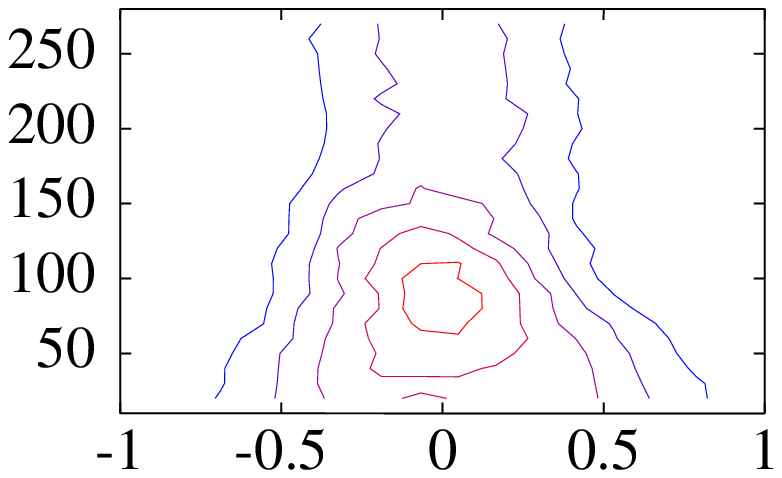}}
   \put(012,022){$1^\text{st}$}
   \put(030,022){\makebox[10mm][r]{\footnotesize\color{red}500}}
   \put(030,019){\makebox[10mm][r]{\footnotesize\color{blue}100}}
   \put(036,000){%
      \includegraphics[bb= 90 80 327 245,clip,width=40mm]{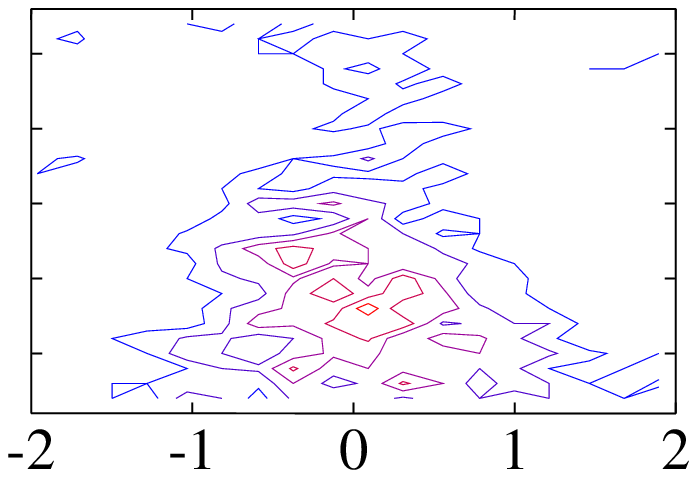}}
   \put(046,022){$2^\text{nd}$}
   \put(064,022){\makebox[10mm][r]{\footnotesize\color{red}25}}
   \put(064,019){\makebox[10mm][r]{\footnotesize\color{blue}5}}
   \put(000,015){\rotatebox[origin=t]{90}{\small $T$ (K)}}
   \put(022,000){\small$q_x$ (\AA$^{-1}$)}
   \put(056,000){\small$q_x$ (\AA$^{-1}$)}
            \end{picture}
            %%%%%%%%%%%%%%%%%%%%%%%%%%%%%%%%%%%%%%%%%%%%%%%%
            \begin{picture}(80,57)
   \put(000,051){(c)}
   \put(029,008){\line(0,1){44}}
   \put(030,050){$T_\text{sc}$}
   \put(048.5,008){\line(0,1){33}}
   \put(049.5,039){$T_\text{mag}$}
   \put(004,002){\includegraphics[width=78mm]{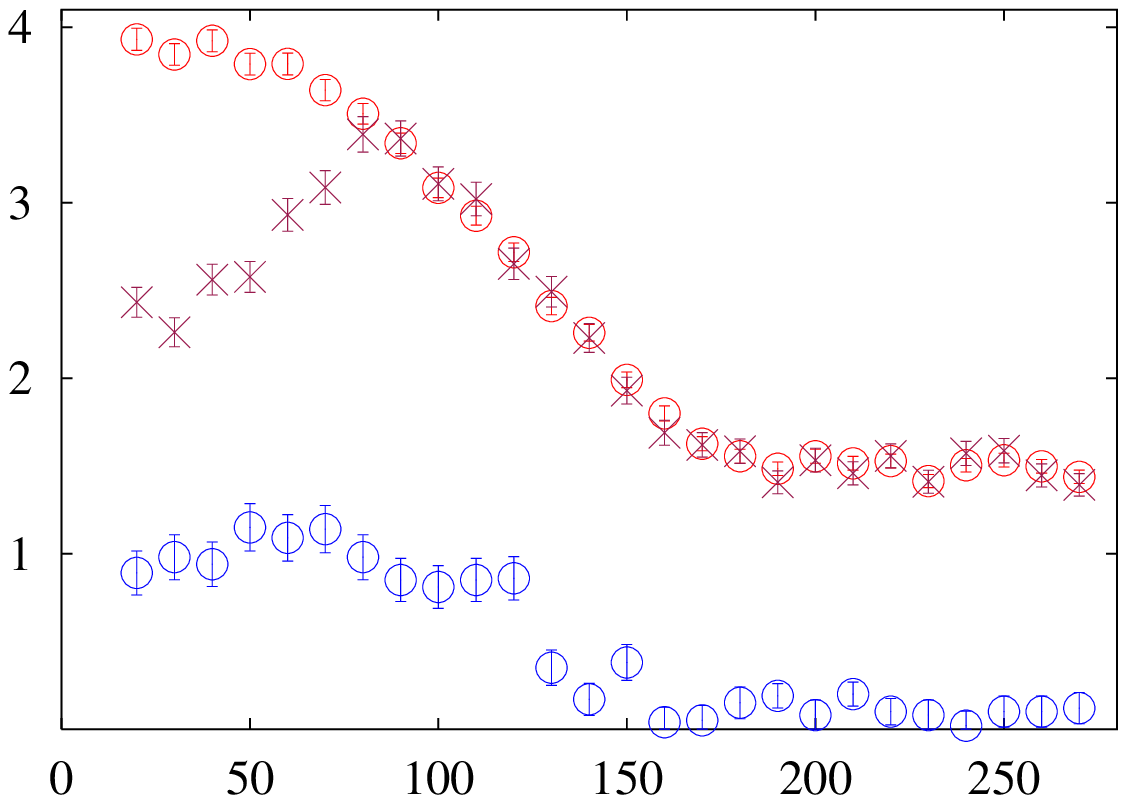}}
   \put(002,022){\small\rotatebox{90}{Intensity $/1000$}}
   \put(040,000){\small $T$ (K)}
   \put(053,029){\small\color{red} $1^\text{st}$ Bragg peak}
   \put(053,014){\small\color{blue} $2^\text{nd}$ Bragg peak}
   \put(063,011){\small\color{blue} $\times 10$}
            \end{picture}
            %%%%%%%%%%%%%%%%%%%%%%%%%%%%%%%%%%%%%%%%%%%%%%%%
            \caption{ \label{sam1}
  (a) Specular reflectivity of sample 1 at 200,
   170, 150, 120, 100, 70, 50 and 15\,K,
   for $H_\text{ext} = 100\,\text{Oe}$
   (field cooled). Curves are
   offset for clarity.  Bragg peaks are marked by arrows.
  (b) $I(q_x,T)$ map for the 1st (\emph{left}) and 2nd
   (\emph{right}) Bragg peaks.
  (c) $T$ dependence
   of the 1st ({\em red}\/) and 2nd ({\em blue}\/) Bragg peak intensities,
   integrated over $q_x$ ({\LARGE$\circ$}) and  at $q_x=0$ ($\times$,
   scaled by 6).
            }
            %%%%%%%%%%%%%%%%%%%%%%%%%%%%%%%%%%%%%%%%%%%%%%%%
           \end{figure}

Figure \ref{sam1}(a) displays unpolarized neutron reflectivity
curves taken under specular condition. The intensity of the 1st
Bragg peak (averaged over the neutron spin states) exhibits a
sizable increase below $T_\textrm{mag}$. This shows that the
magnetic potential $V_\text{mag}(z)$ enhances the contrast between
the YBCO and LCMO layers, as expected due to the onset of magnetic
order in LCMO. However, the $1:1$ ratio of the layer thicknesses
dictates that the even-order Bragg peaks should not be observable
if the FM order parameter is either absent, or spatially uniform
and confined to the LCMO layer. This condition is fulfilled for
the curves at $T > T_\text{mag}$ where the 2nd Bragg peak at $q_z
\sim 0.07\,\text{\AA}^{-1}$ (as marked by the shaded area) is
absent to within the noise level. Below $T_\text{mag}$, however,
the 2nd Bragg peak suddenly appears.
The magnetic origin of the 2nd Bragg peak (confirmed by
the polarized-beam data below) is indicative of a substantial
difference between the spatial profiles of the nuclear and
magnetic potentials.

A second kind of remarkable anomaly occurs in the vicinity of the
SC transition. It is best seen in the off-specular rocking scans
at the Bragg positions, as shown in Figures \ref{sam1}(b) and (c).
The off-specular reflectivity is sensitive to a momentum transfer
parallel to the plane of the multilayer ($q_x$) and thus provides
information on \textit{in-plane} correlation of the nuclear and
magnetic profiles. The left panel of Fig. \ref{sam1}(b) shows that
the off-specular scattering is weak and nearly temperature
independent for $T > T_\text{mag}$. Such diffuse scattering is
characteristic of uncorrelated in-plane roughness of the nuclear
potential.

Below $T_\text{sc}$, however, a pronounced broadening occurs, and
weight is transferred from the specular to the off-specular part.
This is readily visible in Fig. \ref{sam1}(c) where the specular
intensity (\textit{crosses}) of the 1st Bragg exhibits a strong
decrease, whereas the intensity integrated along $q_x$
(\textit{circles}) increases. The observed trend is indicative of
a profound SC-induced increase in the magnetic roughness. A
corresponding trend is observed for the 2nd Bragg peak as shown in
the right panel of Fig. \ref{sam1}(b). For the 1st Bragg peak we
estimate a change of the full-width at half maximum $\Delta q_x$
from $0.6\cdot10^{-4}\,\text{\AA}^{-1}$ (close to the instrumental
resolution) at 75\,K to about $1.2\cdot10^{-4}\,\text{\AA}^{-1}$
at 15\,K. This translates into a change of the characteristic
%length scale
magnetic domain size from more than $15\,\mu\text{m}$ at $T \simeq
T_\text{sc}$ to about $7\,\mu\text{m}$ at $T\ll T_\text{sc}$.
These pronounced SC-induced changes of the in-plane component of
the magnetic profile are suggestive of a sizable proximity
coupling of the SC and FM order parameters. SQUID magnetization
data (not shown) indicate that the FM magnetic moments are
oriented parallel to the layers of our superlattices. Orbital
effects of the magnetic field in the SC layers are therefore
expected to be weak, and the dominant interaction is the magnetic
exchange coupling. This introduces a spin-splitting of the
electronic states and reduces the SC condensation energy. In
return, the development of the SC order parameter
favors the formation of FM domain boundaries where the
pair-breaking is substantially reduced \cite{buzd88}. The
anomalous decrease in the size of the FM domains therefore
provides a clear indication for a strong proximity coupling
between the SC and FM order parameters. 
A spontaneous vortex phase (due to a minor perpendicular component of the FM moments) 
in the SC layers or the presence
of an unconventional SC order parameter with a spin-triplet
component could also contribute to the off-specular signal
\cite{buzd03}. These scenarios could be tested by mapping out the
off-specular signal in further experiments with an improved
signal-to-noise ratio.

\begin{figure}
            \unitlength1mm
            %%%%%%%%%%%%%%%%%%%%%%%%%%%%%%%%%%%%%%%%%%%%%%%%
            \begin{picture}(80,55)
   \put(000,051){(a)}
   %\put(043,009){{\color{gray}\rule{12mm}{16mm}}}
   \put(004,003){\includegraphics[width=78mm]{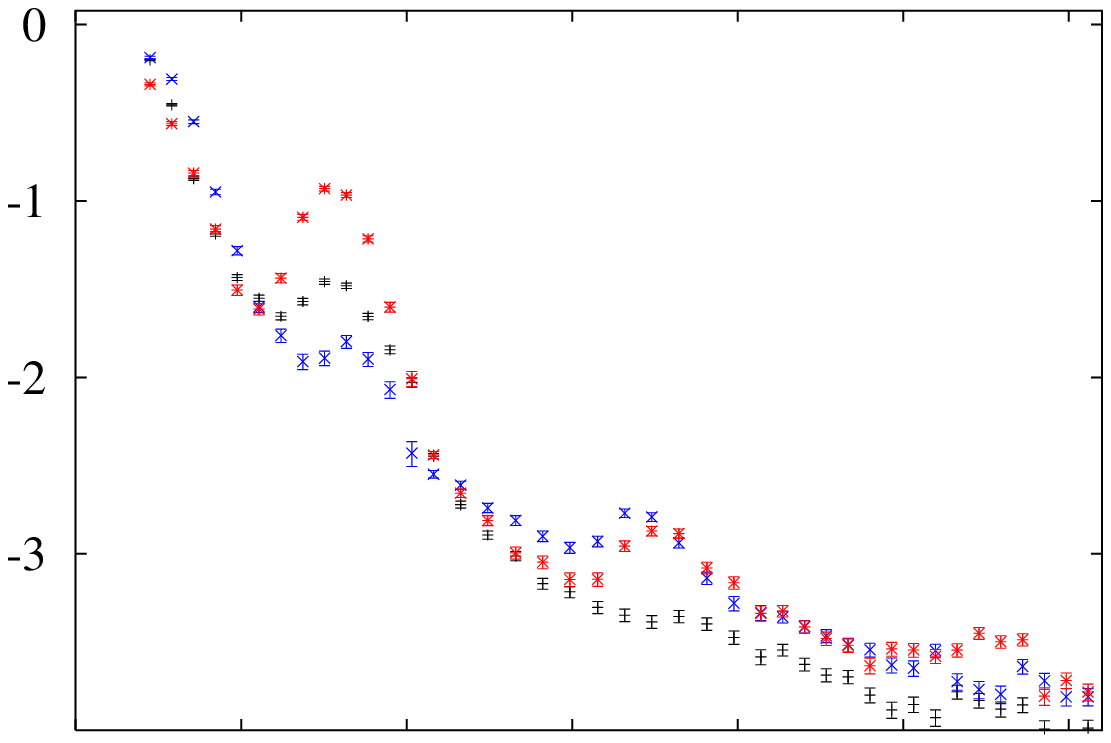}}
   \put(002,028){\rotatebox[origin=t]{90}{$\log_{10}[R(q_z)]$}}
   \put(029,045){$\downarrow  1^\text{st}$}
   \put(048,027){$\downarrow  2^\text{nd}$}
   \put(070,018){$\downarrow  3^\text{rd}$}
   \put(015,023){$\color{red} R^{|->}_{15\,\text{K}}$}
   \put(015,017){$\color{blue} R^{|+>}_{15\,\text{K}}$}
   \put(015,012){$R_{200\,\text{K}}$}
            \end{picture}
            %%%%%%%%%%%%%%%%%%%%%%%%%%%%%%%%%%%%%%%%%%%%%%%%
            \begin{picture}(80,46)
   \put(000,051){(b)}
   %\put(043,009){{\color{gray}\rule{12mm}{16mm}}}
   \put(004,003){\includegraphics[width=78mm]{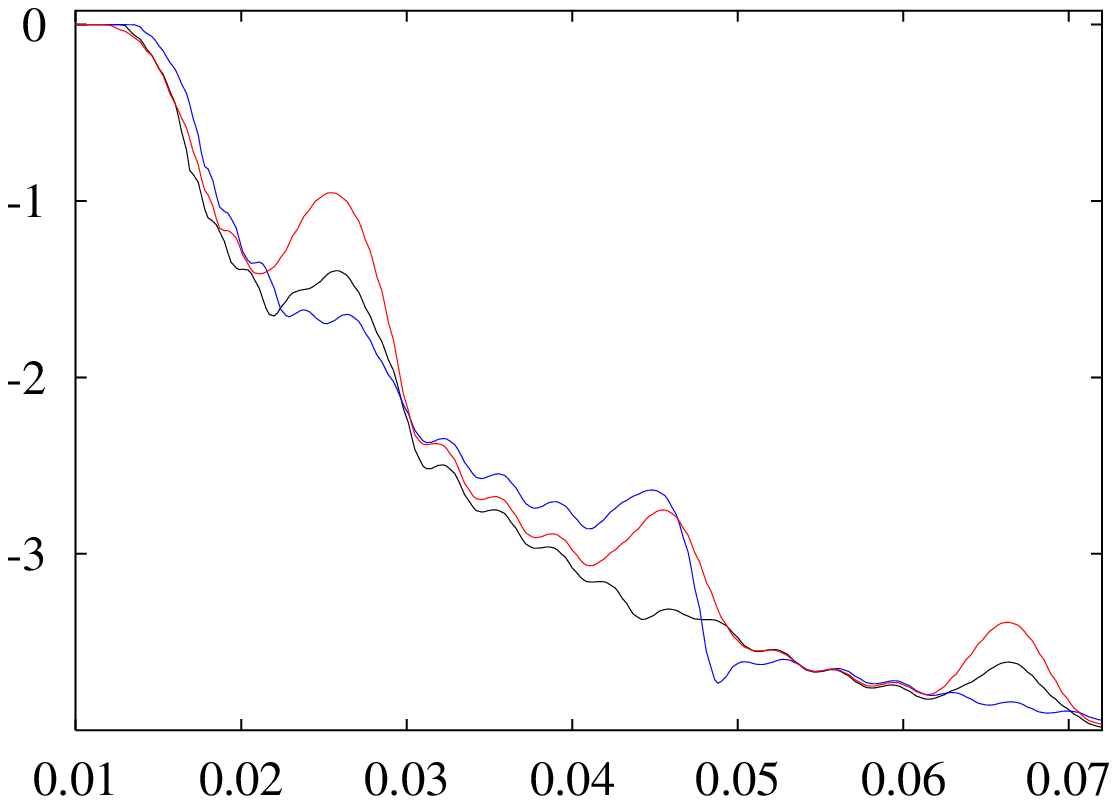}}
   \put(002,028){\rotatebox[origin=t]{90}{$\log_{10}[R(q_z)]$}}
   \put(042,000){$q_z$ (\AA$^{-1}$)}
   \put(029,045){$\downarrow  1^\text{st}$}
   \put(048,027){$\downarrow  2^\text{nd}$}
   \put(070,018){$\downarrow  3^\text{rd}$}
   \put(015,023){$\color{red} R^{|->}_\text{calc}$}
   \put(015,017){$\color{blue} R^{|+>}_\text{calc}$}
   \put(015,012){$R_\text{calc, non-magnetic}$}
            \end{picture}
            %%%%%%%%%%%%%%%%%%%%%%%%%%%%%%%%%%%%%%%%%%%%%%%%
            \unitlength0.8mm
            \begin{picture}(100,67)
   \put(000,056){(c)}
   \put(036,008.7){{\color{gray}\rule{15.45mm}{41.15mm}}}
   \put(076,008.7){{\color{gray}\rule{15.45mm}{41.15mm}}}
   \put(011,002){\includegraphics[width=38.0mm,height=50mm]{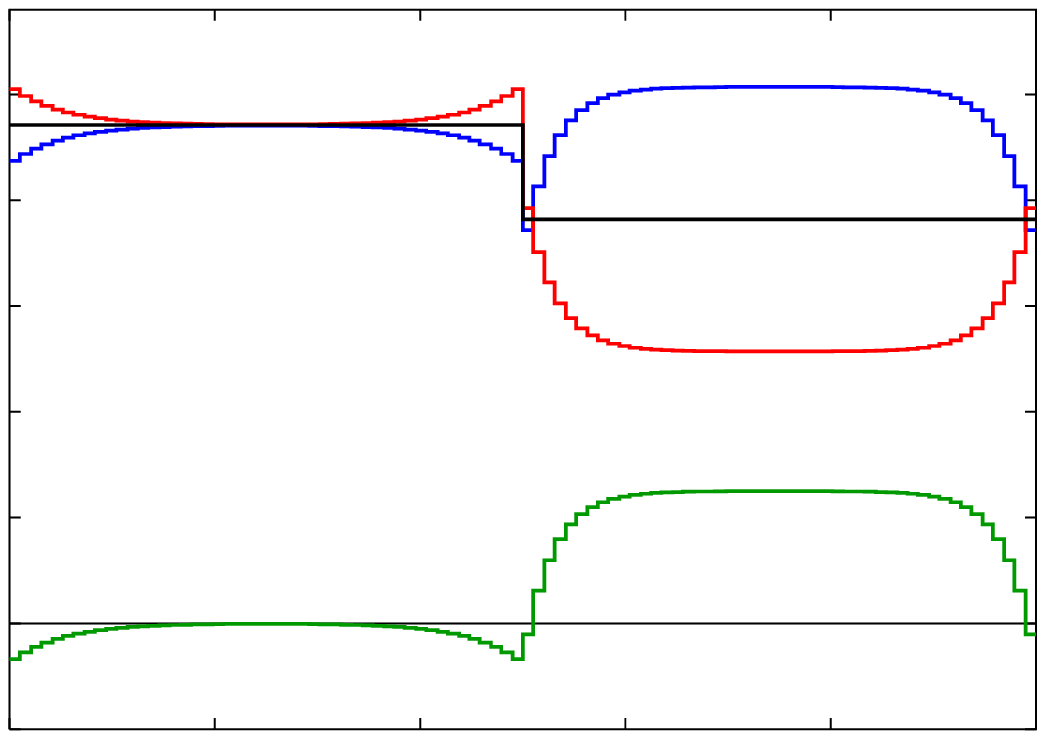}}
   \put(051,002){\includegraphics[width=38.0mm,height=50mm]{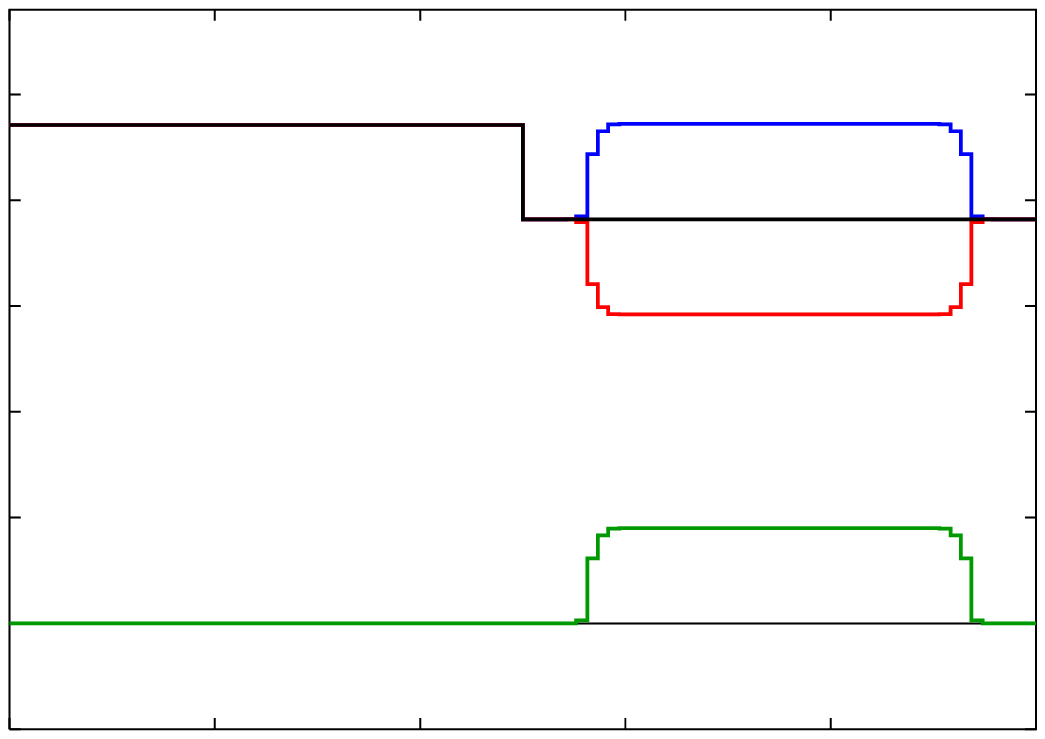}}
   \put(005,032){\rotatebox[origin=t]{90}{$\frac{2\pi}{\lambda^2}\,
              \delta(z)\cdot 10^6$}}
   \put(018,056){\small model 1}
   \put(058,056){\small model 2}
   \put(012,053.9){\small 5}
   \put(012,045.8){\small 4}
   \put(012,037.9){\small 3}
   \put(012,030.3){\small 2}
   \put(012,023){\small 1}
   \put(012,015.6){\small 0}
   \put(017,004){\small 0}
   \put(034,004){\small 0.5}
   \put(054,004){\small 1}
   \put(057,004){\small 0}
   \put(074,004){\small 0.5}
   \put(094,004){\small 1}
   \put(040,000){$z$ within one period}
   \put(020,030){\small YBCO}
   \put(041,030){\small LCMO}
   \put(022,048){\small $\text{nuc}$}
   \put(022,013){\color{darkgreen}\small $\text{mag}$}
   \put(043,052){\color{blue}\small $_{|+>}$}
   \put(043,038){\color{red}\small $_{|->}$}
   \put(060,030){\small YBCO}
   \put(081,030){\small LCMO}
            \end{picture}
            %%%%%%%%%%%%%%%%%%%%%%%%%%%%%%%%%%%%%%%%%%%%%%%%
            \caption{\label{sam2}%
   (a)
    Polarized specular reflectivity of
    sample 2 at 200 and 15\,K, for
    $H_\text{ext} = 100\,\text{Oe}$ (field cooled).
   (b)
    Simulated reflectivity curves (model 1) and
   (c)
    model potentials that reproduce the experimental data.
    \textit{Left}: inverse proximity effect (model 1) and
    \textit{right}: ``dead layer'' (model 2).
    $\delta(z) \propto V(z)$ is the deviation of
    the refractive index from 1, $\lambda$ is the neutron wavelength.
            }
            %%%%%%%%%%%%%%%%%%%%%%%%%%%%%%%%%%%%%%%%%%%%%%%%
           \end{figure}

We now describe a quantitative analysis of the magnetization
profile perpendicular to the layers. We tested numerous models
with the EDXR code \cite{edxr} that allows one to compare the
calculated reflectivity curves with the experimental ones. In
order to separate structural and magnetic contributions and to
determine the quality of the interface, the nuclear contribution
$V_\text{nuc}(z)$ was determined from the curves at $T >
T_\text{mag}$. The obtained individual layer thicknesses are
98\,\AA\ and 160\,\AA\, respectively, for samples 1 and 2. The
density for LCMO was reduced by 2\% with respect to the bulk
value.
The interface was described by a roughness of $\sigma \approx
5\,\text{\AA}$, which testifies to the high quality of our
superlattices. It is well known that neutron reflectivity curves
lack phase information and thus cannot be uniquely assigned to a
particular density or magnetization profile. Nevertheless, we are
able to identify only two possible solutions. The main challenge
in selecting an appropriate magnetization profile is to reproduce
the well-defined 1st structural Bragg peak, the magnetically
induced 2nd Bragg peak, and the low intensity of the 3d structural
Bragg peak. An extra constraint is imposed by the marked
differences in polarized up-spin and down-spin reflectivities as
shown in Fig. \ref{sam2}(a). While a wide variety of models are
able to reproduce the 1st Bragg peak, the presence of the 2nd peak
demonstrates that the magnetic potential $V_\text{mag}(z)$ cannot
simply follow the block-like nuclear profile. In the calculations,
the homogeneous magnetization of YBCO and LCMO layers was replaced
by 48 slices of equal thickness with individually varying
magnetization. Based on extensive computer simulations, we were
able to exclude several physically meaningful models: (i) An
antiferromagnetic coupling between the ferromagnetic layers would
lead to a doubled period and hence to additional Bragg peaks, at
$q_z = 0.022\,\text{\AA}$ and at $q_z = 0.053\,\text{\AA}$ which
are absent in the reflectivity curves. (ii) A magnetic roughness
of any length-scale only leads to a faster decay of the
reflectivity but not to a second Bragg peak. (iii) A conventional
magnetic proximity effect where the magnetization exhibits an
exponential decay into the SC layer also fails.

In the following we discuss the only two successful models for
which the magnetization profiles are illustrated in Fig.
\ref{sam2}(c). Model 1 (\textit{left panel}) contains a sizable
magnetic moment within the YBCO layer that couples
antiferromagnetically to the one in LCMO (inverse proximity
effect). Notably, the antiparallel alignment is essential to
reproduce the observed 2nd Bragg peak positions and intensities.
Model 2 (\textit{right panel}) assumes a ``dead'' region with no
net magnetic moment (either paramagnetic or antiferromagnetic)
within the LCMO layer. The resulting fits of the polarized
reflectivities using model 1 are shown in Fig. \ref{sam2}(b). The
values of the magnetic induction thus obtained are $B_\text{YBCO}
= 0.6\,\text{T}$ and $B_\text{LCMO} = 0.9\,\text{T}$ (sample 1)
and $B_\text{YBCO} = 0.6\,\text{T}$ and $B_\text{LCMO} =
1.4\,\text{T}$ (sample 2). The
%typical length scale
penetration depth of the interfacial magnetization profile is of
the order of $20\,\text{\AA}$ in the YBCO layer, and
$10\,\text{\AA}$ in the LCMO layer. Similar fits were obtained
with model 2 (not shown) assuming a thickness of the ``dead
layer'' of $\approx 20\,\text{\AA}$ and magnetic induction $B$ of
$1.1\,\text{T}$ (sample 1) and $1.5\,\text{T}$ (sample 2).

           %================================================
           %================================================

While the model calculations do not allow one to differentiate
between these two cases, we argue that the anomalous $T-$dependence
in the SC state is a clear indication for a strong mutual
interaction between the SC and FM order parameters and thus
strongly favors model 1, which implies close contact between the
SC and FM order parameters. First, we point out that the combined
results of x-ray diffraction, electron microscopy \cite{habe} and
in particular neutron reflectivity curves for $T > T_\text{mag}$
testify to the high quality of the interfaces with practically
absent inter-growth and small overall structural roughness of the
order of 5\,\AA. The magnetic profiles deduced from our neutron
reflectivity data are therefore not merely due to interfacial
disorder like inter-growth or inter-diffusion. One might argue
that the magnetically ``dead layer" in LCMO, the centerpiece of
model 2, could arise from interfacial strain or charge transfer
across the interface. According to the phase diagram of LCMO, this
could introduce an insulating layer with antiferromagnetic order.
However, such a ``dead layer" would also efficiently reduce the
exchange coupling between the SC and FM order parameters. Since
the FM moments are oriented parallel to the layers (as
demonstrated by SQUID magnetization data) and the magnetic
penetration depth of the superconductor is very large for this
field orientation, the electromagnetic coupling between FM and SC
is also expected to be negligible.

In contrast, model 1 describes a situation where the SC and FM
order parameters are in close contact and thus are likely to
experience a strong mutual interaction. Interestingly, the
essential feature of model 1, that is, a thin layer on the SC side
which has a net magnetic moment oriented antiparallel to the one
in the FM layer, has recently been proposed theoretically
\cite{berg}. According to this theory, the unusual magnetization
profile near the interface originates from Cooper pairs that have
a finite overlap with both the FM and the SC layers.
Heuristically, the preferential spin alignment of one electron in
the FM layer leads to an antiparallel spin orientation of the
second electron of the spin-singlet pair that resides in the SC
layer. While phase-coherent Cooper pairs exist only in the SC
state, \textit{i.e.}\/ for $T<T_\textrm{sc}$, similar arguments
apply for other kinds of itinerant spin-singlet pairs.
The existence of spin-singlet pairs at elevated temperatures $T>
T_{\textrm{sc}}$ has  been indeed proposed in the context
of the unusual normal state electronic properties of the cuprate
HTSC (so-called pseudogap phenomenon) \cite{gap}. Our data are
also consistent with recent macroscopic magnetization measurements
suggesting an antiferromagnetic component of the magnetization
profile at the YBCO/LCMO interface \cite{haberkorn}. Underdoped
cuprates are known to be susceptible to antiferromagnetic order,
and a staggered magnetization profile whose amplitude decreases as
a function of distance from the interface would generate a net
magnetization in YBCO, as observed.

In summary, our neutron reflectometry measurements on high-quality
$\mathrm{YBa_2Cu_3O_7/La_{2/3}Ca_{1/3}MnO_3}$ multilayers have
revealed detailed, microscopic information about the magnetization
profile as a function of in-plane and out-of-plane wave vectors.
The pronounced modification of this profile at the superconducting
transition  indicates a significant proximity coupling
between SC and FM order parameters, likely due to exchange
interactions. The findings are discussed in terms of the recently predicted 
inverse proximity effect \cite{berg}.

% % % % % % % % % % % % % % % % % % % % % % % % % % % % % % % % % % % % %
We acknowledge M.\ Wolff for support on ADAM at
ILL, France. This work was partly performed at Morpheus at SINQ, Paul
Scherrer Institute, Switzerland.
%%%%%%%%%%%%%%%%%%%%%%%%%%%%%%%%%%%%%%%%%%%%%%%%%%%%%%%%%%%%%%%%%%%%%%%%%%

%%%%%%%%%%%%%%%%%%%%%%%%%%%%%%%%%%%%%%%%%%%%%%%%%%%%%%%%%%%%%%%%%%%%%%%%%%
%%%%%%%%%%%%%%%%%%%%%%%%%%%%%%%%%%%%%%%%%%%%%%%%%%%%%%%%%%%%%%%%%%%%%%%%%%

\end{document}